\pdfoutput=1

\documentclass[prl,showpacs,twocolumn,superscriptaddress,nofootinbib]{revtex4-1}

\usepackage{amsmath,amsfonts,amssymb,amstext}
\usepackage[dvipdfm]{graphicx}
\usepackage[colorlinks,linkcolor=blue]{hyperref}
\newcommand{\bra}[1]{\ensuremath{\langle #1 \vert}}
\newcommand{\ket}[1]{\ensuremath{\vert #1 \rangle}}
\newcommand{\braket}[2]{\ensuremath{\langle #1 \vert #2 \rangle}}
\newcommand{\ave}[1]{\langle #1 \rangle}

\begin{document}
\title{Enhanced squeezing of a collective spin via control of its qudit subsystems}
\date{\today}

\author{Leigh M. Norris}
\email{norrisl@unm.edu}
\affiliation{Center for Quantum Information and Control (CQuIC)}
\affiliation{Department of Physics and Astronomy, University of New Mexico}

\author{Collin M. Trail}
\affiliation{Institute for Quantum Information Science, University of Calgary}

\author{Poul S. Jessen}
\affiliation{Center for Quantum Information and Control (CQuIC)}
\affiliation{College of Optical Sciences and Department of Physics, University of Arizona}

\author{Ivan H. Deutsch}
\affiliation{Center for Quantum Information and Control (CQuIC)}
\affiliation{Department of Physics and Astronomy, University of New Mexico}

\begin{abstract}
Unitary control of qudits can improve the collective spin squeezing of an atomic ensemble.  Preparing the atoms in a state with large quantum fluctuations in magnetization strengthens the entangling Faraday interaction. The resulting increase in interatomic entanglement can be converted into metrologically useful spin squeezing. Further control can squeeze the internal atomic spin without compromising entanglement, providing an overall multiplicative factor in the collective squeezing.  We model the effects of optical pumping and study the tradeoffs between enhanced entanglement and decoherence.  For realistic parameters we see improvements of $\sim$10 dB.
\end{abstract}

\maketitle

Large atomic ensembles interacting with optical fields show great promise as platforms for quantum metrology \cite{Polzik2009, Vuletic2010} and quantum memory \cite{DLCZ, Fleischhauer2002, Polzik2004, Kuzmich2004, Kimble2008}. An important benchmark for such protocols is spin squeezing arising from entanglement between atoms \cite{Kuzmich2000}. To create these states, modes of the optical field can be used as a quantum data bus, inducing entanglement through their collective coupling to all atoms. Spin-squeezed states have direct application in atomic clocks \cite{wineland}, magnetometry \cite{Mitchell2010, Budker2012}, and continuous variable quantum information processing \cite{ContinuousVar}. 

In most studies of such applications, the atoms are treated as qubits, with only two internal levels participating in the interaction, e.g., the ``clock states'' of $^{133}$Cs. However, the hyperfine ground manifold of cesium, and most atoms used in such experiments, are naturally $d>2$ dimensional qudits with a richer structure~\cite{Toth2011} that one can potentially harness to improve the squeezing protocol. In this paper we show that unitary control of internal atomic magnetic sublevels, or ``qudit control", along with collective control via the atom-light interface can strongly enhance our ability to create nonclassical states of the ensemble. The key idea is that the entanglement generated between the atoms and photons depends strongly on the internal state of the atoms.  Through local unitary control of the atomic spin qudit, we can increase this entanglement and then map it into a useful form. We benchmark this enhancement by calculating the achievable spin squeezing that is metrologically relevant for precision magnetometry.

The goal of spin squeezing is to reduce the variance of the collective spin in some direction below the standard quantum limit. For $N_A$ atomic spins with hyperfine spin quantum number $f$, defining $\hat{F}_z =\sum_i \hat{f}_z^{(i)}$, the collective spin variance of a state symmetric under interchange of atoms is 
\begin{equation}
\Delta F_z^2 = N_A(N_A-1)  \ave{\Delta \hat{f}_z^{(i)} \Delta \hat{f}_z^{(j)}}_{i\neq j}+N_A\Delta f_z^2. 
\end{equation}
Entanglement between atomic spins can make the first term negative, reducing the collective spin variance. Qudit control enables us to apply an identical unitary transformation to each atom in the ensemble.  State preparation using such control can strengthen the coupling between the light and the ensemble, translating into increased interatomic entanglement. In addition, qudit control can also squeeze the internal spin uncertainty, $\Delta f_z^2$, below the standard quantum limit, an effect not possible for spin-1/2 atoms~\cite{Chaudhury07, Fernholz08}.  We will show how to achieve this without compromising the squeezing arising from entanglement~\cite{MultilevelHP}.  

We study here the Faraday interaction, which entangles the collective atomic spin with the Stokes vector describing the polarization state of a quantized optical probe mode, according to the unitary operator $\hat{U}=\exp\{-i\chi \hat{F}_z \hat{S}_3\}$ \cite{InterfaceHammerer, LongPaper}.  The 3-component of the quantized Stokes vector,  $\hat{S}_3$, corresponds to the difference in the number of right and left circularly polarized photons. For atoms with spin $f>1/2$, the atomic-spin/photonic-Stokes interaction contains an additional birefringent effect that couples the atomic alignment to the $\hat{S}_2$ and $\hat{S}_1$ Stokes vector components.  These can be removed by applying sequences of  alternating orthogonally polarized probe pulses \cite{MMDynamicalDecoupling} or a large bias field along the $z$-axis, which effectively averages the relative direction between the mean atomic spin and light polarization.  

Consider the action of the Faraday operator on an initial product state of the atoms and light,
\begin{equation}
\hat{U}\ket{\Psi_A}\otimes\ket{\Phi_L}= \sum_{F,M_z} C_{F,M_z} \ket{F,M_z}\otimes  e^{iM_z \chi \hat{S}_3} \ket{\Phi_L},
\end{equation}
where we have decomposed the atomic state in terms of the basis of collective angular momentum states $\ket{F,M_z}$ (Dicke states).  The amount of entanglement generated between the light and atoms is determined by the {\em distinguishability} of the rotated states $e^{iM_z \chi \hat{S}_3} \ket{\Phi_L}$ for different values of the collective magnetization, $M_z$.  Initial states of the atomic ensemble with greater quantum uncertainty in $M_z$ (referred to as ``projection noise"  in quantum metrology) translate into a larger variation in Faraday rotation angles.  When this variation becomes observable in a shot-noise limited measurement of the probe polarization, there is significant entanglement between the atoms and photons (Fig. 1).  It is this atom-light entanglement that is ultimately converted into entanglement of the atoms with one another, and thus is the resource thats acts to squeeze the collective spin.

We quantify the entanglement generated by the Faraday interaction with a ``measurement strength'', $\xi$,  given by the ratio of the collective atomic projection noise (PN) to the measurement resolution as determined by the probe shot noise (SN). \cite{LongPaper}.  For an initial product state of $N_A$ atoms, the projection-noise uncertainty is   $\left(\Delta F_z^2\right)_{PN} =N_A \Delta f_z^2$, and the shot noise for $N_L$ photons  in a time $\tau$ is $\left(\Delta F_z^2\right)_{SN} = (\chi^2 N_L)^{-1}$ \cite{LongPaper}, so,
\begin{equation}
\xi =\frac{\left( \Delta F_z^2\right)_{PN}}{\left(\Delta F_z^2\right)_{SN}}=\chi^2 N_L N_A \Delta f^2_z=\frac{1}{9} (\gamma_s \tau) OD \frac{\Delta f^2_z}{f^2}. 
\label{xi}
\end{equation}
Here, $\chi$ is the Faraday rotation angle per unit angular momentum for a spin-$f$ atom interacting with a laser beam detuned from a $S_{1/2}\rightarrow P_{J}$  transition, $OD$ is the optical density for a unit oscillator strength, and $\gamma_s$ is the photon scattering rate per atom, as defined in~\cite{LongPaper}. 

 \begin{figure}
[t]\resizebox{8.7cm}{!}
{\includegraphics[scale=0.5]{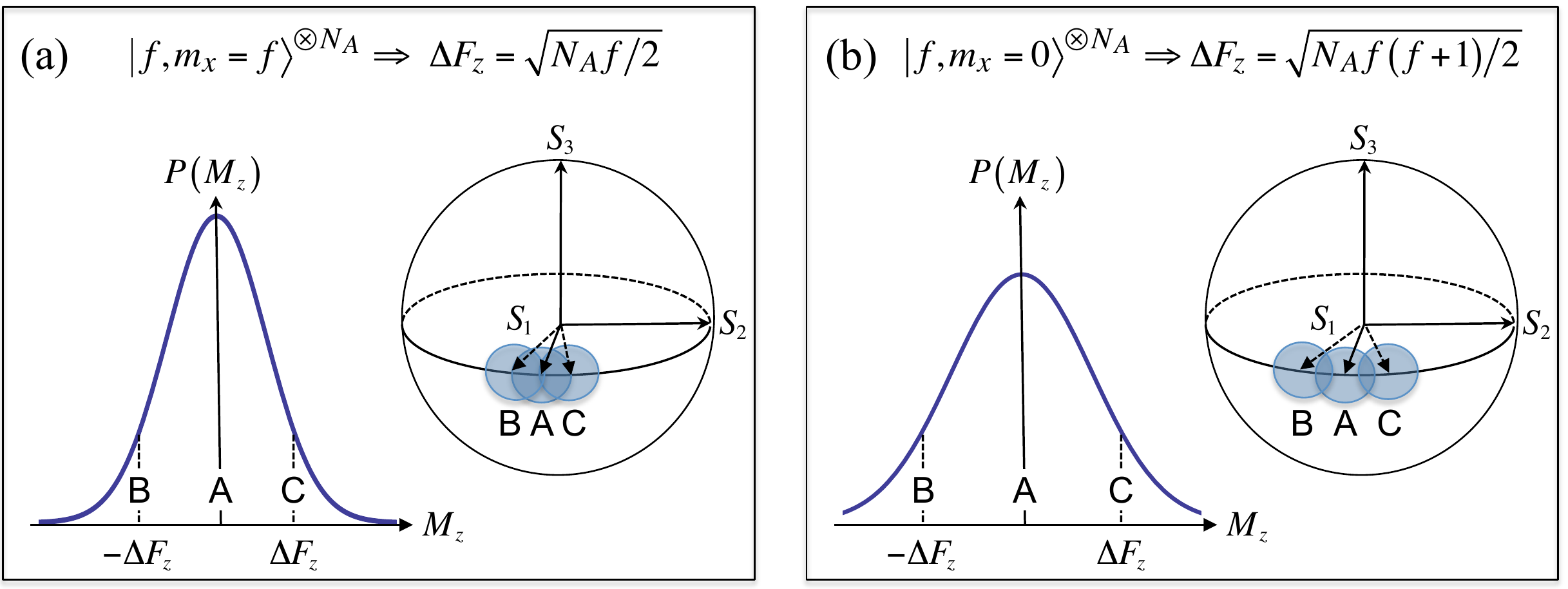}}
\caption{\label{fig:figure1} The Faraday interaction rotates the Stokes vector about the $S_3$ axis by an amount proportional to the collective magnetization $M_z$ of the atoms.  For a large ensemble of $N_A$ atoms in identical states, the probability distribution $P(M_z)$ is Gaussian with variance $\Delta M_z^2 = N_A \Delta f_z^2$.  The light is initially in a coherent state along $S_1$ with shot noise fluctuations along $S_2$ and $S_3$.  The quantum uncertainty in $M_z$ leads to a spread in Faraday rotation angles. Shown in schematic are the spread in rotations for atoms initially in (a) a spin coherent state  along $x$, $\ket{f,m_x=f}^{\otimes N_a}$,  and (b) the state $\ket{f,m_x=0}^{\otimes N_a}$.  The increased quantum uncertainty in $M_z$ (``projection noise'') translates into increased entanglement between atoms and photons as the corresponding rotations become more distinguishable for a given probe shot noise.}
\end{figure}

Equation (\ref{xi}) immediately suggests how qudit control can be employed to increase the entanglement generated by the Faraday interaction. Our choice of the initial (fiducial) state should maximize the  uncertainty in collective magnetization so the possible values of $M_z$ are as distinguishable as possible.  Formally, we see this in the spin squeezing produced by a quantum nondemolition (QND) measurement of the collective $\hat{F}_z$ mediated by the Faraday interaction. Consider an ensemble of atoms identically prepared in a desired fiducial state $\ket{\Psi_A}_{in}=\ket{\uparrow}^{\otimes N_A}$.  For small Faraday rotation angles, we can linearize the interaction, $\hat{U}\approx 1-i\chi \hat{F}_z \hat{S}_3$.  The coherent dynamics then couples the fiducial internal state to one other state $\hat{f}_z \ket{\uparrow} = c \ket{\downarrow}$, where we have chosen the phase of $\ket{\downarrow}$ so that $c$ is real.  Under the assumption $\bra{\uparrow}\hat{f}_z\ket{\uparrow}=0$, it follows that $\braket{\downarrow}{ \uparrow} =0$ and $c=\sqrt{\Delta \hat{f}^2_{z_\uparrow}}$, so  $\hat{f}_z \approx \sqrt{\Delta \hat{f}^2_{z_\uparrow}} \left(\ket{\uparrow}\bra{\downarrow}+\ket{\downarrow}\bra{\uparrow}\right)$.  After a Faraday interaction with $N_L$ linearly polarized photons in the state $\ket{\Phi_L}$, the light and atoms become entangled, and quantum backaction disturbs the atomic state conditioned on a measurement of the light. 

In the QND protocol we probe the atoms with light polarized along the $S_1$ direction of the Poincar\'{e} sphere and measure the $S_2$ component at the output.  We consider the case when the meter reads $S_2=0$, for which $\bra{0_L}\hat{S}_3\ket{\Phi_L}=0$, and $\bra{0_L}\hat{S}_3^2\ket{\Phi_L}=\braket{0_L}{\Phi_L}(N_L/2 )$ . Other values of $S_2$ yield a displaced squeezed state.  The post-measurement state of the atoms, to lowest nonvanishing order in the measurement strength, is
\begin{eqnarray}
&\ket{\Psi_A}_{out}=\bra{0_L} e^{i \chi \hat{F}_z \hat{S}_3}\ket{\Phi_L} \ket{\Psi_A}_{in} \\ \nonumber
	&  \approx \braket{0_L}{\Phi_L}\left\{ \ket{\Psi_A}_{in} - \frac{\chi^2 N_L}{4} \hat{F}_z^2 \ket{\Psi_A}_{in}\right\} \\ \nonumber
	& \approx \ket{\uparrow}^{\otimes N_A}- \frac{\xi}{4 N_A}\sum\limits_{i\neq j} \ket{\downarrow_i \downarrow_j} \ket{\uparrow}^{\otimes (N_A-2)}_{\neq i,j},
\label{pairwise}
\end{eqnarray}
where we have renormalized at the last step.  The pairwise entanglement between fiducial and coupled states, $\ket{\uparrow \uparrow}-  \frac{\xi}{4 N_A}\ket{\downarrow\downarrow}$, is the essence of collective spin squeezing.  From Eq. (\ref{xi}),  $\xi \propto \Delta f_z^2$, and thus increased projection noise in the fiducial state leads to enhanced atom-atom entanglement.  

To illustrate this formalism, consider first a preparation in a spin coherent state ({\em SCS}) for arbitrary $f \ge 1/2$.  The fiducial state is polarized in the $x$-direction, $\ket{\uparrow_{SCS}} = \ket{f,m_x=f}$.  The coupled state is $\ket{\downarrow_{SCS}}= \ket{f,m_x=f-1}$, since $\hat{f}_z \ket{\uparrow_{SCS}}= \sqrt{f/2} \ket{\downarrow_{SCS}}$ and $\Delta f^2_{z,SCS} = f/2$.  The measurement strength for the {\em SCS} fiducial state is  $\xi_{SCS}= OD \gamma_s \tau/(18f)$.  Enhancement of the measurement strength is achieved by using fiducial states with larger projection noise.  The maximum value is attained by the ``cat-state'' preparation of the internal spin, $\ket{\uparrow_{cat}} \equiv \left(\ket{f,m_{z}=f}+\ket{f,m_{z}=-f}\right)/\sqrt{2}$ with $\Delta f^2_{z,cat} = f^2$.  It follows that the orthogonal coupled state is $\ket{\downarrow_{cat}} =\left(\ket{f,m_{z}=f}-\ket{f,m_{z}=-f}\right)/\sqrt{2}$, since $\hat{f}_z\ket{\uparrow_{cat}} =f \ket{\downarrow_{cat}}$.  Thus, $\xi_{cat} = OD \gamma_s \tau/9$, a factor of $2f$ improvement over the {\em SCS}.  As discussed below, the cat-state preparation is more easily damaged by decoherence due to photon scattering.  We therefore consider a third potential preparation, $\ket{\uparrow_{0_x}} =\ket{f,m_{x}=0}$, for which $\Delta f^2_{z,0_x} = f(f+1)/2$, a factor $f+1$ enhancement over the {\em SCS}.  For this choice, the coupled state is $\ket{\downarrow_{0_x}} =  \left(\ket{f,m_{x}=1}+\ket{f,m_{x}=-1}\right)/\sqrt{2}$, since $f_z \ket{\uparrow_{0_x}}=\sqrt{f(f+1)/2}\ket{\downarrow_{0_x}}$.

To study the enhanced squeezing, we employ a multilevel Holstein-Primakov approximation (HPA) \cite{MultilevelHP}.  Writing the collective spin in second quantization, $\hat{F}_z = \sqrt{\Delta f^2_{z_\uparrow}} \left(\hat{a}^{\dag}_{\uparrow} \hat{a}_{\downarrow} + \hat{a}^{\dag}_{\downarrow} \hat{a}_{\uparrow}\right)$, and linearizing about the mean field,  $\hat{a}_{\uparrow}\approx \sqrt{N_{\uparrow}}\approx\sqrt{N_A}$, we have $\hat{F}_z \approx \sqrt{2N_A\Delta f^2_{z_\uparrow}}\hat{X}_{\downarrow}$, where $(X_\downarrow , P_\downarrow )$ are the quadratures associated with quantum fluctuations of $\hat{a}_\downarrow$.  The initial state of the ensemble, with all atoms in the fiducial state $\ket{\uparrow}$, corresponds to the vacuum. 

Entangled spin states generated by the Faraday interaction correspond to quadrature squeezing in the $(X_\downarrow , P_\downarrow )$ plane, but not necessarily spin-squeezing.  As our metric, we use the spin squeezing parameter relevant for magnetometry, $\zeta = 2f N_A \Delta F_\perp^2/\left|\ave{\hat{F}_x}\right|^2$ \cite{wineland}, where $F_\perp$ is the squeezed component.  While the $\ket{cat}^{\otimes N_A}$ and $\ket{0_x}^{\otimes N_A}$ preparations enhance interatomic entanglement, the resulting state is not spin-squeezed;  for these states, $\ave{\hat{F}_x}=0$.  However, given the capability to perform arbitrary unitary transformations on the internal hyperfine levels, the entanglement can be made metrologically useful by mapping the embedded qubit according to the isometry $\hat{U} = \ket{f,m_x=f}\bra{\uparrow}+ \ket{f,m_x=f-1}\bra{\downarrow}$.  We locally map the internal fiducial and coupled states for any preparation to those associated with the {\em SCS}; the squeezing created in a quadrature of the $\left(X_\downarrow, P_\downarrow \right)$ plane is transferred to reduced fluctuations around the  {\em SCS}. After this mapping, the squeezing of $X_\downarrow$ is equivalent to spin squeezing, $\zeta=2\Delta X_\downarrow^2$. The protocol is thus as follows: prepare all atoms in a desired $\ket{\uparrow}$, thereby increasing the distinguishability of the collective magnetization via the Faraday interaction.  The laser field then mediates enhanced entanglement between $\ket{\uparrow \uparrow}$ and $\ket{\downarrow\downarrow}$, and this entanglement is mapped back to the {\em SCS} where it is metrologically useful.  

As a final step, we employ qudit control to squeeze the internal $f>1/2$ spins of each atom~\cite{Chaudhury07, Fernholz08}.  The key idea here is to do so in a manner that multiplies the final metrologically relevant squeezing rather than reducing the squeezing already incurred via atom-atom entanglement~\cite{MultilevelHP}. We consider the family of squeezed Yurke-like states for integer $f$~\cite{Yurke1986, wineland} 
\begin{equation}
\ket{yur} \equiv  \frac{1}{\sqrt{2}} \sin\alpha \ket{1_z} +\cos\alpha  \ket{0_z} +\frac{1}{\sqrt{2}} \sin\alpha \ket{-1_z},
\end{equation}
written in the basis $\ket{m_z} = \ket{f,m_z}$, for which $\zeta_Y =( 1+f)^{-1}(\cos \alpha)^{-2}$. We can combine the effects of entanglement-induced squeezing with internal squeezing by the appropriate mapping of the embedded qubit. We choose as our final fiducial state $\ket{\uparrow_Y}=\ket{yur}$ and the coupled state  $\ket{\downarrow_Y}$ according to $\hat{f}_z\ket{\uparrow_Y}=\sqrt{\Delta \hat{f}^2_{z}}\ket{\downarrow_Y}$, where $\ket{\downarrow_Y} =\left(\ket{1_z}-\ket{-1_z}\right)/\sqrt{2}$.  When entanglement is created  between the states of in the initial preparation, e.g., $\{\ket{\uparrow_{cat}\uparrow_{cat}}\ket{\downarrow_{cat}\downarrow_{cat}}\}$, these correlations can be transferred to Yurke-squeezed pairs, $\{\ket{\uparrow_Y \uparrow_Y},\ket{\downarrow_Y \downarrow_Y}\}$, using internal control acting locally on each atom. We thereby map the squeezed quadrature fluctuations from the $(X_\downarrow, P_\downarrow)_{cat}$ plane to the $(X_\downarrow, P_\downarrow)_{Y}$ plane.  The metrologically relevant spin squeezing under this mapping becomes, $\zeta = ( 1+f)^{-1}(\cos \alpha)^{-2} \, 2 (\Delta X_\downarrow^2)_{cat}$, enhanced over the coherent state by both  the increased measurement strength and a factor that grows with increasing atomic spin $f$.

To study the effectiveness of qudit control in enhancing spin squeezing, we apply the techniques described here to the protocol detailed in~\cite{DoublePass}, which utilizes coherent feedback to create an optimal two-axis countertwisting unitary \cite{KitagawaUeda} on the collective spin.  In the absence of decoherence, within the HPA, one can show that the squeezing parameter is $\zeta = e^{-\xi}(1+f)^{-1}\left(\cos\alpha\right)^{-2}$, including the final Yurke map.  Enhancing $\xi$ by increasing initial projection noise fluctuations has an substantial effect on the squeezing that is achievable in a short time. 

\begin{figure}
[t]\resizebox{8.7cm}{!}
{\includegraphics{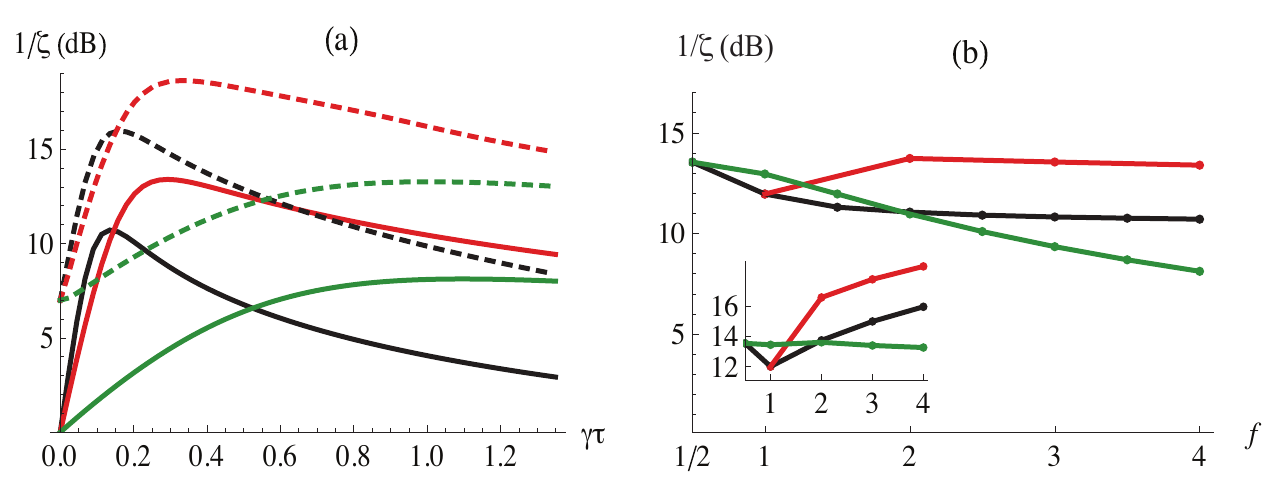}}
\caption{\label{fig:figure2} (Color online) Numerically calculated squeezing (dB).  (a) Squeezing as a function of time in units of the scattering rate $\gamma$ for an ensemble of $f=4$ cesium atoms with OD=300, and a probe detuned $\Delta=10^3\Gamma = 33$ GHz from the D2 line, for three state preparations: {\em SCS} (green), {\em cat} (black), $0_x$ (red).  The dashed curves include the final squeezing of the internal spin. (b) Peak squeezing of the three state preparations as a function of $f$.  The inset includes the final Yurke mapping. }  
\end{figure}

The choice of fiducial state and the ultimate performance of this protocol is set by the tradeoffs between enhanced squeezing and increased noise due to decoherence.  Here, the fundamental source of decoherence is optical pumping due to photon scattering.  We consider alkali-metal atoms  in the electronic-ground $S_{1/2}$ hyperfine manifold $f=i+1/2$, for nuclear spin $i$, and linearly polarized light, detuned far compared with the excited $P_{3/2}$ hyperfine splitting.  Optical pumping gives rise to three processes: (i) ``spin-flips'' between the fiducial state and the coupled state; (ii) loss to the other hyperfine manifold; (iii) pumping to other magnetic substates within the manifold $f$, but outside the embedded qubit subspace. Each of these processes leads to a reduction in the pairwise entanglement that contributes to spin squeezing. Processes (i) and (iii)  also lead to additional noise, arising from statistical mixtures of atoms in different $\ket{f,m_f}$.  In the case of process (iii), however, this excess noise can be removed by application of a microwave pulse that transfers these atoms to  $f=i-1/2$ after the Faraday interaction. If this is done, processes (ii) and (iii) are equivalent. We cannot use this approach for atoms that have undergone process (i), as these are still in the encoded qubit subspace, with correlations relevant to spin squeezing. Each state preparation is subject to different rates of processes (i), (ii), and (iii). The {\em SCS} preparation has the lowest rate of spin flips, $\gamma_{flip}^{SCS}= \gamma/(12f)$, followed by the  $0_x$ with $\gamma_{flip}^{0_x} =\gamma (f+1)/(18f)$ and lastly the $cat$ with  $\gamma_{flip}^{cat}= \gamma/9$. Since the total rate of optical pumping events is $2\gamma/9$ for all state preparations, states with fewer spin flips will have greater loss.  In this sense, the cat state is most susceptible to decoherence.

An additional subtlety is when both processes (i) and (iii) occur but are not perfectly distinguishable.  In this case there is a ``transfer of coherence'' \cite{CohenTannoudji1977} that can reduce the amount of additional noise added in a spin flip event. For example, in the {\em SCS} case, a photon scattering event can pump an atom from $\ket{f, m_x=f} \rightarrow \ket{f, m_x=f-1}$ (process (i))  or  $\ket{f, m_x=f-1} \rightarrow \ket{f, m_x=f-2}$ (process (iii)).  While it may seem advantageous to remove the excess noise in the additional state $\ket{f, m_x=f-2}$, the transfer of coherences and the resulting squeezing make it beneficial to retain it.  These transferred coherences reduce the noise arising from spin flips $\ket{f, m_x=f}\rightarrow \ket{f, m_x=f-1}$.  A similar effect is seen in the $0_x$ preparation.  The cat-state preparation has no useful transfer of coherence and thus is more fragile to decoherence.

To calculate the squeezing produced in this protocol, we use the covariance matrix formulation as discussed in~\cite{MadsenMolmer}; details will be presented elsewhere. We track the fluctuations in the quadratures $(X_\downarrow, P_\downarrow)$ and their correlations.  For a given fiducial state preparation, the final squeezing in these quadratures is mapped to those around {\em SCS}, and from this we determine the metrologically relevant $\zeta$. When the transfer of coherences are important, we retain a third state that is mapped to $\ket{f, m_x=f-2}$.  Finally, we can also include an additional step and squeeze the internal spin via the Yurke-state mapping.  To account for the transfer of coherences, the third state is mapped to $\ket{\overline{yur}} = \cos\alpha \left(\ket{1_z} +\ket{-1_z} \right)/\sqrt{2}+\sin\alpha \ket{0_z}$.  

Figure 2a shows the squeezing as a function of time for $f=4$, for a nominal unit-oscillator-strength optical density $OD=300$, and a detuning from the D2 line $6S_{1/2}\rightarrow 6P_{3/2}$, $\Delta=10^3\Gamma$.  The initial rate of squeezing is largest for the cat-state, but the $0_x$ state ultimately produces the most squeezing because of its greater robustness to decoherence.  Internal spin squeezing adds about 7 dB of squeezing for this large spin.  

In Fig. 2b, we study the peak squeezing produced by these different protocols as a function of $f$, showing the subtle tradeoffs between enhanced coupling and fragility to decoherence.  In the absence of the Yurke-state map, for $f=1$ the {\em SCS} performs best because of the transfer of coherence. For larger $f$, however, the state's robustness to decoherence does not compensate for the reduction in $\xi_{SCS}$.  For the $0_x$ preparation, the peak squeezing is largest, and can outperform spin-1/2 ensembles.  Including the final internal-spin squeezing, a larger $f$ yields more overall squeezing for the {\em cat} and $0_x$ preparations. 

The protocol presented here has important implications for quantum-limited metrology.  While we have used  magnetometry as our benchmark, the entanglement produced by the Faraday interaction with a well-chosen fiducial state can be transferred to correlations between atoms in the clock state, providing potentially larger squeezing for Ramsey interferometry than other protocols~\cite{Polzik2009, Vuletic2010}.  The ultimate limit of any protocol will depend on the delicate tradeoffs between enhanced measurement strengths and decoherence.  While we have demonstrated the potential use of this control in the examples of the {\em cat} and $0_x$ preparations, in future work we will seek the optimal fiducial state for a given noise model.   

This work was supported by NSF Grants PHY-0903953, 0969371, 0969997,  AFOSR Grant FA95501110156, and by PIMS and NSERC.
 
\bibliography{QuditReferences}
\end{document}